\documentclass[useAMS]{mn2e}
\usepackage{graphicx}
\usepackage{longtable}

\title[Star-forming regions in the western tidal tail of NGC 2782]{NGC 2782: a merger remnant with young stars in its gaseous tidal tail\thanks{Based on observations obtained at the Gemini Observatory, which is operated by the Association of Universities for Research in Astronomy, Inc., under a cooperative agreement with the NSF on behalf of the Gemini partnership: the National Science Foundation (United States), the Science and Technology Facilities Council (United Kingdom), the National Research Council (Canada), CONICYT (Chile), the Australian Research Council (Australia), Minist\'erio da Ci\^encia e Tecnologia (Brazil) and Ministerio de Ciencia, Tecnolog\'ia e Innovaci\'on Productiva (Argentina) -- Observing run: GN-2009B-Q-113.}}

\author[S. Torres-Flores et al.]
{
\parbox[t]{\textwidth} {S. Torres-Flores$^{1,2,3}$\thanks{Current address. E-mail: storres@dfuls.cl}, C. Mendes de Oliveira$^{1}$, D. F. de Mello$^{4,5}$, S. Scarano Jr$^{1}$ \& F. Urrutia-Viscarra$^{1}$}
\vspace*{6pt}\\
$^{1}$Departamento de Astronomia, Instituto de Astronomia, Geof\'isica e Ci\^encias Atmosf\'ericas da USP,\\ Rua do Mat\~ao 1226, Cidade Universit\'aria, 05508-090, S\~ao Paulo, Brazil\\
$^{2}$Laboratoire d'Astrophysique de Marseille, OAMP, Universit\'e de Provence \& CNRS,\\ 38 rue F. Joliot--Curie, 13388 Marseille, Cedex 13, France\\
$^{3}$Departamento de F\'isica, Universidad de La Serena, Av. Cisternas 1200 Norte, La Serena, Chile \\
$^{4}$Observational Cosmology Laboratory, Code 665, Goddard Space Flight Center, Greenbelt, MD 20771, USA\\
$^{5}$Catholic University of America, Washington, DC 20064, USA
}

\begin{document}

\date{}

\pagerange{\pageref{firstpage}--\pageref{lastpage}} \pubyear{2010}

\maketitle

\label{firstpage}

\begin{abstract}
We have searched for young star-forming regions around the merger remnant
NGC 2782. By using \textit{GALEX} FUV and NUV imaging and HI data we
found seven UV sources, located at distances
greater than 26 kpc from the center of NGC 2782, and coinciding with
its western HI tidal tail. These regions were resolved in several smaller systems when Gemini/GMOS r-band images were used. 
We compared the observed colors to stellar population synthesis models and we found that these objects have ages of $\sim$1 to 11
Myr and masses ranging from 10$^{3.9}$ to 10$^{4.6}$M$_{\odot}$. By using Gemini/GMOS spectroscopic data we confirm
memberships and derive high metallicities for three of the young
regions in the tail (12+log(O/H)=8.74$\pm$0.20, 8.81$\pm$0.20 and 8.78$\pm$0.20). 
These metallicities are similar to the value presented by the nuclear region of NGC 2782 
and also similar to the value presented for an object located close to the main body of
NGC 2782. The high metallicities measured for
the star-forming regions in the gaseous tidal tail of NGC 2782 could be
explained if they were formed out of highly enriched gas which was once
expelled from the center of the merging galaxies when the system collided. An additional possibility is that the
tail has been a nursery of a few generations of young stellar systems
which ultimately polluted this medium with metals, further enriching the
already pre-enriched gas ejected to the tail when the galaxies collided.
\end{abstract}


\section{Introduction}

The intergalactic medium (IGM) of interacting galaxies and the tidal tails of mergers have shown to commonly be the birthplace of a number of new stellar systems. These systems span the range from small clusters (Knierman et al .2003, Tran et al. 2003) and intergalactic H\,{\sc ii} regions (IH\,{\sc ii}, e. g. Mendes de Oliveira et al. 2004) to Tidal Dwarf Galaxies (TDGs, Mirabel et al. 1992, Duc \& Mirabel 1998, Weilbacher et al. 2000, Weilbacher et al. 2003, Bornaud et al. 2004, Mundell et al. 2004, Mendes de Oliveira et al. 2006). 

Since the total luminosity of these young stellar systems is mostly dominated by the ultraviolet (UV) ionizing radiation coming from massive stars, the use of the \textit{Galaxy Evolution Explorer} (\textit{GALEX}) satellite has become an important tool to study the star formation processes and the stellar populations of these systems (Neff et al. 2005, Hancock et al. 2007, de Mello, Torres-Flores, Mendes de Oliveira  2008, Smith et al. 2008, Torres-Flores et al. 2009, Hancock et al. 2009, Boquien et al. 2009, Smith et al. 2010, Boquien et al. 2010).

Another crucial ingredient for the study of the stellar populations and formation processes of young stellar systems is the knowledge of their element abundances. This can, in fact, constrain the formation scenario of these stellar associations. For example, Weilbacher et al. (2003) studied the oxygen abundance of several knots in tidal features which had been classified as TDG candidates. These authors found metallicities similar to the values found in the outer disk of the spirals from which these knots were formed, which suggests that these objects were born from a pre-enriched material. In a lower mass regime, Mendes de Oliveira et al. (2004) studied spectroscopically four H\,{\sc ii} regions in the intergalactic medium of the compact group HCG 92. These authors found high metallicities for these regions, also suggesting that they were formed from a pre-enriched material.

In this context, we have searched for star-forming regions in the merger remnant NGC 2782, which include spectroscopic confirmation for three of the sources. This allows us to study the formation of stellar clusters, the enrichment of the intergalactic medium and to attempt to identify the evolutionary stage of the merger.

NGC 2782 (Arp 215) is at a distance of 34 Mpc (Smith 1994). Its heliocentric radial velocity is 2543$\pm$2 km s$^{-1}$ (from NED database). This galaxy shows a prominent tidal tail detected in HI, located to the western side of the object. A second tidal tail formed by a stellar component is seen to the east side of the galaxy. Smith (1994) reports the presence of arcs and ripples in this system. These kinds of signatures are common in colliding and merging galaxies (Schweizer 1980, 1982). Due to all the features listed above, Smith (1994) and Smith et al. (1999) classified this system as a merger remnant. In the tidal tail located to the east side of the galaxy, Yoshida et al. (1994) detected a dwarf galaxy in formation and molecular gas was found by Smith et al. (1999) in this region. No molecular gas was found in the western tidal tail (Smith et al. 1999, Braine et al. 2001). New CO observations of the central region of NGC 2782 are shown in Hunt et al. (2008). In a search for extended ultraviolet disks, Thilker et al. (2007) included the merger remnant NGC 2782 in its sample. These authors report the presence of UV clumps in the northwest tidal tail of NGC 2782, suggesting that star formation is taking place in this tail.

Recently, Werk et al. (2011) studied the oxygen abundances of several outlying H\,{\sc ii} regions in a sample of local galaxies, where NGC 2782 was included. Theses authors concluded that most of their systems have flat metallicity gradients. In \S 4.2 we discuss the results found by Werk et al. with ours.

This paper is organized as follows: In \S 2 and \S 3 we present the data and data analysis. In \S 4 we present the results. In \S 5 we discuss our results and 
in \S 6 we present our main conclusions.

\section{Data}

\subsection{Ultraviolet Data}

The UV \textit{GALEX} images of NGC 2782 have been published by Thilker et al. (2007). In our case, the UV data analyzed in this work were taken from the public archival data of the \textit{GALEX} satellite. From this archive, we used images in the near ultraviolet (NUV $\lambda$$_{\rm eff}$=2271\AA) and 
far ultraviolet (FUV $\lambda$$_{\rm eff}$=1528\AA) bands. In the NUV-band, the exposure time was 2109 sec and in the FUV band, it was 2108 sec. FUV and NUV fluxes were calculated using Morrisey et al. (2005) 
m$_{\lambda}$=-2.5 log[F$_{\lambda}$/a$_{\lambda}$] + b$_{\lambda}$, 
where a$_{FUV}$ = 1.4 $\times$ 10$^{-15}$ erg s$^{-1}$ cm$^{-2}$ \AA$^{-1}$,  
a$_{NUV}$=2.06$\times$ 10$^{-16}$ erg s$^{-1}$ cm$^{-2}$ \AA$^{-1}$, b$_{FUV}$=18.82 and b$_{NUV}$=20.08 for FUV and NUV, respectively. Fluxes were multiplied by the effective filter bandpass ($\Delta$$\lambda$$_{FUV}$=269\AA\, and $\Delta$$\lambda$$_{NUV}$=616\AA) to give units of erg s$^{-1}$ cm$^{-2}$. The \textit{GALEX} fields of view are 1$^{\circ}$.28 and 1$^{\circ}$.24 in FUV and NUV respectively and the pixel scale is 1.5 arcsec pixel$^{-1}$. The images had a resolution (FWHM) of 4.2$\arcsec$ and 5.3$\arcsec$ in FUV and NUV, respectively.

\subsection{r-band and spectroscopic data}

Observations of NGC 2782 were carried out with the Gemini multi object spectrograph (GMOS) at the Gemini north observatory as part of the science program GN-2009B-Q-113. r-band images for two fields in NGC 2782 were observed. Fields were centered on the western gaseous tidal tail (RA=09:13:53.5, DEC=40:09:31.6, J2000) and on the object NGC 2782 (RA=09:14:03.2, DEC=40:06:59.1, J2000). The exposure times for both r-band images were 15 minutes, under a seeing of 0.54" and 0.77", respectively. Zero point calibrations were taken from the GEMINI website. We checked these values using a few bright stars (on the same field) from the \textit{SDSS} database.

Archival GMOS multi slit spectroscopic exposures (3$\times$1200 s) were available for five objects on the western tidal tail of NGC 2782, covering from 3400 to 6100 \AA (grating B600). This Gemini data were originally acquired by Werk et al. (2011), who provide more details on the observations and an independent reduction and analysis (science program GN-2008A-Q-31). In this work, these spectra were reduced using standard routines in IRAF\footnote{IRAF is distributed by the National Optical Astronomy Observatories, which are operated by the Association of Universities for Research in Astronomy, Inc., under cooperative agreement with the National Science Foundation. See http://iraf.noao.edu}. In addition, an \textit{SDSS} spectrum (covering from 3800 to 9229 \AA) was available for one object in the star-forming arc along the inner ripple of NGC 2782. This object, whose coordinates are RA=09:14:02.6 and DEC=+40:06:47.2 (J2000), was already detected in the H$\alpha$ images shown by Hodge \& Kennicutt (1983), Smith (1994, Fig. 6) and Jogee et al. (1998, Fig. 4).

\subsection{Neutral gas}

The calibrated HI data of NGC 2782 was taken from the NED database. This has a spectral resolution of 31.5 km s$^{-1}$. Details about the observation can be found in Smith (1994).

\section{Data Analysis}

\subsection{Selection criteria}

In this work, we are interested in searching for young star-forming regions in the gaseous tidal tail of NGC 2782, therefore no regions on the main body of this galaxy were taken into account. The spectrum of a typical star-forming region is obviously characterized by a strong H$\alpha$ line. If we simulate an instantaneous burst of star formation using the STARBURST99 model (SB99, Leitherer et al. 1999), the lifetime of the H$\alpha$ line, given a fixed mass of 10$^{6}$M$_{\odot}$, is of the order of 10 Myr (for a solar metallicity). This is basically produced by the short lifetime of massive stars (OB stars). When this age is linked to the ultraviolet colors, it results in a color FUV-NUV=-0.10. However, when the metallicity increases (to Z=2.5Z$_{\odot}$, for instance) the models predict an age of 10 Myr for a color FUV-NUV=0.03. In this theoretical scenario, objects having colors bluer than FUV-NUV$\sim$0 have experienced a recent episode of star formation. 

On the other hand, we note what has been measured from observations. The colors FUV-NUV of the objects studied in Mendes de Oliveira et al. (2004), as measured in Torres-Flores et al. (2009) are (in three of the four regions) FUV-NUV$\leq$0. Also, the TDG candidates studied in de Mello et al. (2008a) with confirmed spectroscopy (de Mello et al. 2012, in preparation) have colors FUV-NUV$\leq$0. Similar colors were found in the study of Neff et al. (2005) for the star forming regions detected in four systems with tidal tails. They found three candidates having masses similar to those of confirmed TDGs and colors FUV-NUV of -0.32$\pm$0.40, -0.09$\pm$0.21 and 0.10$\pm$0.23. Therefore, we selected in this study only objects that have FUV-NUV$\leq$0.15 and that are placed within the HI distribution. 

\subsection{Source extraction and photometry}

We detected UV sources using the software SExtractor (SE, Bertin \& Arnouts 1996) over the NUV and FUV images. Due to the large format of the \textit{GALEX} images, star-forming regions were searched for across the whole HI tail. We set the parameter DETECT$_{-}$THRESH to 1.5$\sigma$ over the background in the NUV image. We chose 1.5$\sigma$ since larger $\sigma$ would cause blending with other sources when using SE automatic apertures. At the end, we had a catalog with RA and DEC of each object in the FUV and NUV bands. With this information in hand, we matched the FUV and NUV catalogs using a 3$''$ radius in order to generate a catalogue in which the sources were identified in both UV bands. As we are interested only in young star-forming regions, we note that our sample as a whole is not complete in FUV magnitudes. FUV and NUV magnitudes were estimated inside a fixed aperture of 4$''$ radius, centered on the centroid of the light distribution of each NUV band detection, using the task PHOT in IRAF. In this case, the sky subtraction was done by using a sky annulus, where the parameter ANNULUS and DANNULUS were set to 10 pixels and the sky fitting algorithm was the mode of the values inside the annulus. Fixed FUV and NUV magnitudes were corrected by aperture effects by using the task MKAPFILE in IRAF. In order to do that, we used the catalog of all the sources available in our field, which is provided by the \textit{GALEX} pipeline. In that catalog, we searched for star-like objects with magnitudes between 15$<$NUV,FUV$<$18 (i. e. non-saturated objects). We used a minimum aperture of 4'' and a maximum one of 60''. This give us a correction of FUV=-0.36$\pm$0.02 and NUV=-0.48$\pm$0.03. In the case of the r-band image, the aperture correction was r=-0.03$\pm$0.01. Due to its small value (similar to the error determination in the r-band magnitudes) we do not take the latter correction into account. Magnitudes were corrected for galactic extinction, using A$_{FUV}$=E(B-V)$\times$8.29, A$_{NUV}$=E(B-V)$\times$8.18 (Seibert et al. 2005). We note that due to the \textit{GALEX} spatial resolution, we can not resolve all of our UV detected regions. In most cases, one UV source will most probably be associated with more than one H\,{\sc ii} region. The optical r-band image will help disentangling the multiple sources. Once the UV-emitting sources were identified, we used the r-band image to obtain the magnitudes. For r-band magnitudes, we used a fixed aperture of 4" radius centered on the UV \textit{GALEX} emission. r-band magnitudes were corrected by Galactic extinction using the Savage \& Mathis (1979) extinction law and the E(B-V) given by Schelegel et al. (1998).

\subsection{Ages and masses: photometric estimates}

For each region, ages were estimated from the FUV-NUV and FUV-r observed colors and the models given by SB99. For these models, which were tuned for the \textit{GALEX} and r-band filters, we used an instantaneous burst, a Salpeter initial mass function (IMF, 0.1 to 100 M$_{\odot}$) and a solar metallicity. Models were generated from 1 Myr to 1 Gyr. It is important to note that for low-mass stellar associations, the observed broad-band colors can be affected by an incomplete sampling of the IMF (effect that is not taken into account by the SB99 models, which assume a fully sampled IMF). This fact could produce uncertainties in the dating of low-mass systems (see Cervi\~no et al. 2003, Cervi\~no \& Valls-Gabaud 2003, Hancock et al. 2008, Popescu \& Hanson 2010 and da Silva et al. 2011). In order to know the color excess (E(B-V)) of each star-forming region, two extinction laws have been applied on the theoretical colors, the starburst law given by Calzetti et al. (1994) and a Milky Way extinction law, where the A$_{FUV}$ and A$_{NUV}$ values were taken from Seibert et al. (2005) (8.29 and 8.18, respectively) and A$_{r}$=2.74 was taken from Savage \& Mathis (1979).

One of the main problems in the study of star formation in young clusters is the uncertainty in the value of the internal dust correction. Since there are no infrared data for the clusters, we have estimated ages and extinctions simultaneously by fitting the observed FUV-NUV and FUV-r colors to the theoretical ones through a $\chi^{2}$ minimization calculation. Errors on ages were estimated at a confidence level of 68$\%$, in a similar way as presented by Smith et al. (2008). We note that the ages and E(B-V) used throughout this paper correspond to the values that minimized the $\chi^{2}$ independent if a starburst or MW extinction law was assumed. Errors on E(B-V) are of the order of 0.3 mag.

We have followed the method used in de Mello et al. (2008a) to estimate the mass of our candidates. Basically, we used the ages and the FUV luminosities of each region to estimate the stellar masses with SB99. The stellar masses were obtained from SB99's monochromatic luminosity, L$_{1530}$, for an instantaneous burst, Salpeter IMF (from 0.1 to 100 M$_{\odot}$) and solar metallicity. Since all our candidates present young ages, we note that the luminosities in L$_{1530}$ are dominated by the contribution of massive stars.

\subsection{Spectroscopic ages and metallicities}

In the \textit{SDSS} database we located a calibrated spectrum for one compact object in an arc along a ripple close to the main body of NGC 2782. Due to its large spectral coverage (from 3800 to 9220 \AA) we can correct this spectrum for reddening using the Balmer emission lines. The value for the intrinsic H$\alpha$/H$\beta$ ratio was taken from Brocklehurst (1971) for an T$_{e}$=10000 K and N$_{e}$=100. The spectrum was then corrected using a c$_{H\beta}$=0.85, using the Cardelli, Clayton \& Mathis (1989) Milky Way extinction law.

GEMINI/GMOS archival spectroscopic data were available for five objects in the gaseous tidal tail of NGC 2782 (data recently used by Werk et al. 2011), three of which coincided with some of the regions measured by us in this study (regions 5, 6 and 7. See \S 4). We used the task EMSAO of the package RVSAO in IRAF to estimate the radial velocity of the objects. Galactic extinction was calculated using the model by Amores \& Lepine (2005), resulting in a E(B-V)=0.08. The intrinsic reddening of each region was assumed to be that derived in \S3.3 (see Table \ref{table1}), given the large uncertainties in the flux estimation of the H$\gamma$ line. Galactic and intrinsic corrections were performed using the IDL code FM\_UNRED by Fitzpatrick (1999).
 
We estimated the equivalent width of the H$\beta$ line for the three regions with GEMINI/GMOS spectroscopic data. Using the observed and synthetic EW(H$\beta$) derived from SB99, we estimated the age of each region, assuming a single stellar population. We also measured the [OII] $\lambda$3726$\textrm{\AA}$, [OIII] $\lambda$4959$\textrm{\AA}$ and [OIII] $\lambda$5007$\textrm{\AA}$ line fluxes relative to the H$\beta$ line. Using these values we estimated the oxygen abundance using the R$_{23}$ method defined by Edmunds \& Pagel (1984, as shown in Fig. 2 of Torres-Peimbert, Peimbert \& Fierro 1989). Although the R$_{23}$ is a well studied calibrator that depends on the fine-tuning between oxygen abundance and electron temperature, it has the disadvantage that it is double valued, presenting an upper and lower branch i. e. for a given value of R$_{23}$, there are two values for the metallicity (see Figure 9 in McGaugh 1991). This can be explained because at low metallicities, the cooling of the H\,{\sc ii} region is dominated by collisionally excited Lyman $\alpha$ emission. While the metallicity increases, the contribution of oxygen to cooling is higher. Optical forbidden lines are very important at values of 12+log(O/H)$\sim$8.4. Beyond that, infrared fine-structure lines dominate the cooling (McGaugh 1991). One way to solve the double valued problem would be by using the ratio between [NII] and [OII], however, for none of the regions in the tail we have spectra with enough dynamic range to see these two lines. Further discussion on this point is given in \S 4.2.

We also estimated the oxygen abundance using the results found by Pilyugin (2001a), who used an empirical approach, which is optimized for high metallicity H\,{\sc ii} regions. In spite of the method shown by Pilyugin (2001a) being optimized for high metallicity H\,{\sc ii} regions, van Zee et al. (2006) found that when this method is compared with direct abundances calculations, the residuals of that comparison seem to be correlated with the ionization parameter of the H\,{\sc ii} regions. van Zee et al. (2006) found that this systematic trend does not exist in the case of the semi-empirical method shown in McGaugh (1991). For this reason, we have also used this latter method to estimate the oxygen abundance of the star-forming regions to compare the results with the other methods mentioned above.

In the case of the \textit{SDSS} spectrum, which does not have the [OII] line, we estimated the oxygen abundance by using the N2 method (Denicol\'o, Terlevich \& Terlevich 2002). 

\section{Results}

In the following we list the main results derived from the photometry and spectroscopy for the star-forming regions detected in NGC 2782. Photometric results are summarized in Table \ref{table1}, while spectroscopic results are listed in Table \ref{table1b}.

\begin{figure*}
\includegraphics[width=\textwidth]{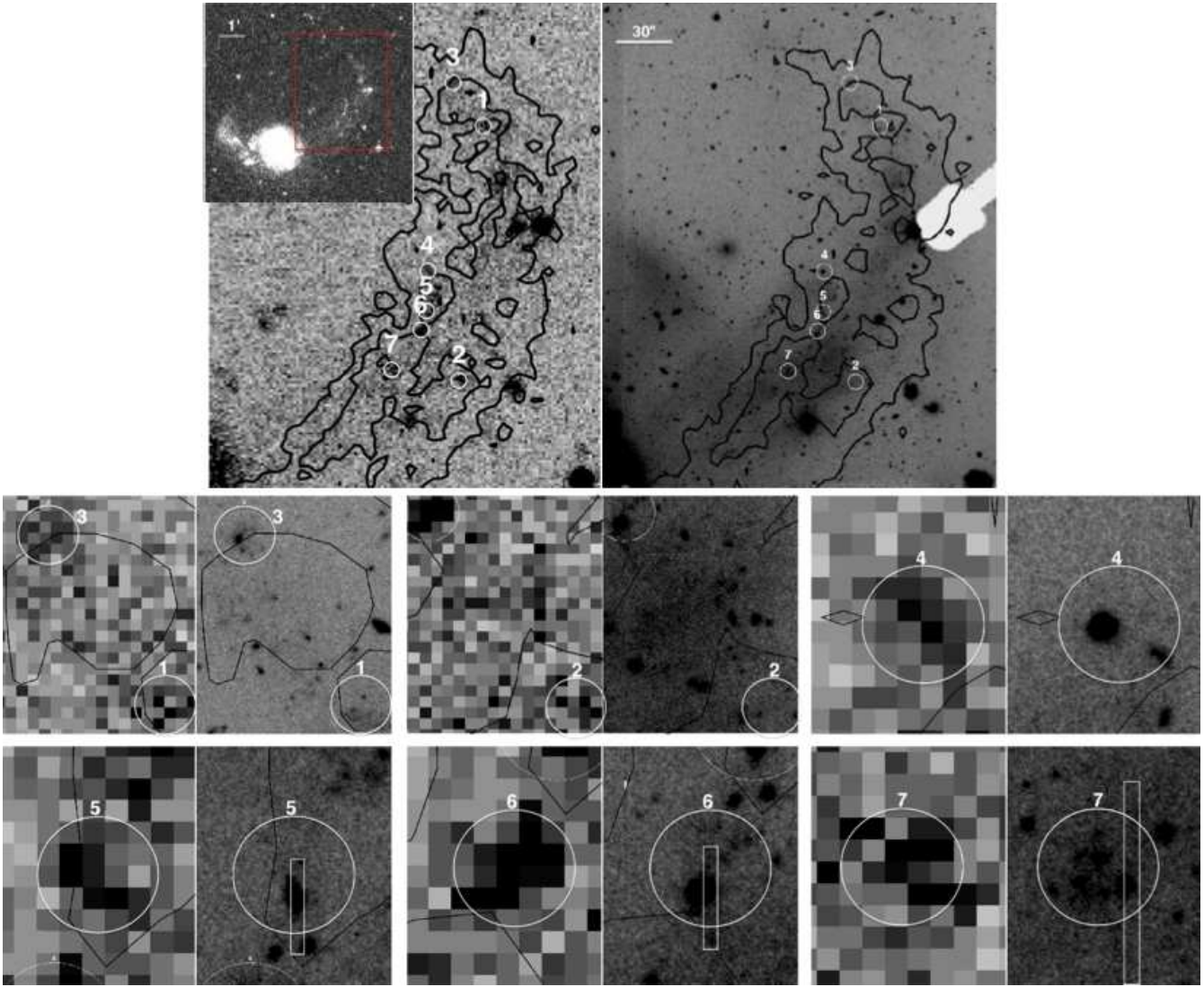}
\caption{Top left: NUV-band image of the HI tidal tail of NGC 2782. A NUV-band image of the entire target is shown in the upper left corner. Top right: r-band image of the tail. Numbered circles (4\arcsec radius) show the detected regions. Contours represent the HI distribution taken from Smith (1994). Bottom images: Close-up of the detected regions (left side: NUV-band image. Right side: r-band image). The GEMINI r-band image resolved the UV detections in several smaller stellar clusters, as exemplified in the lowest 6 panels. White rectangles over regions five, six and seven indicate, approximately, the position of the slit in the spectroscopic observation.}
\label{fig1}
\end{figure*}

\subsection{Main photometric properties of the star-forming regions}

\begin{figure}
\includegraphics[width=\columnwidth]{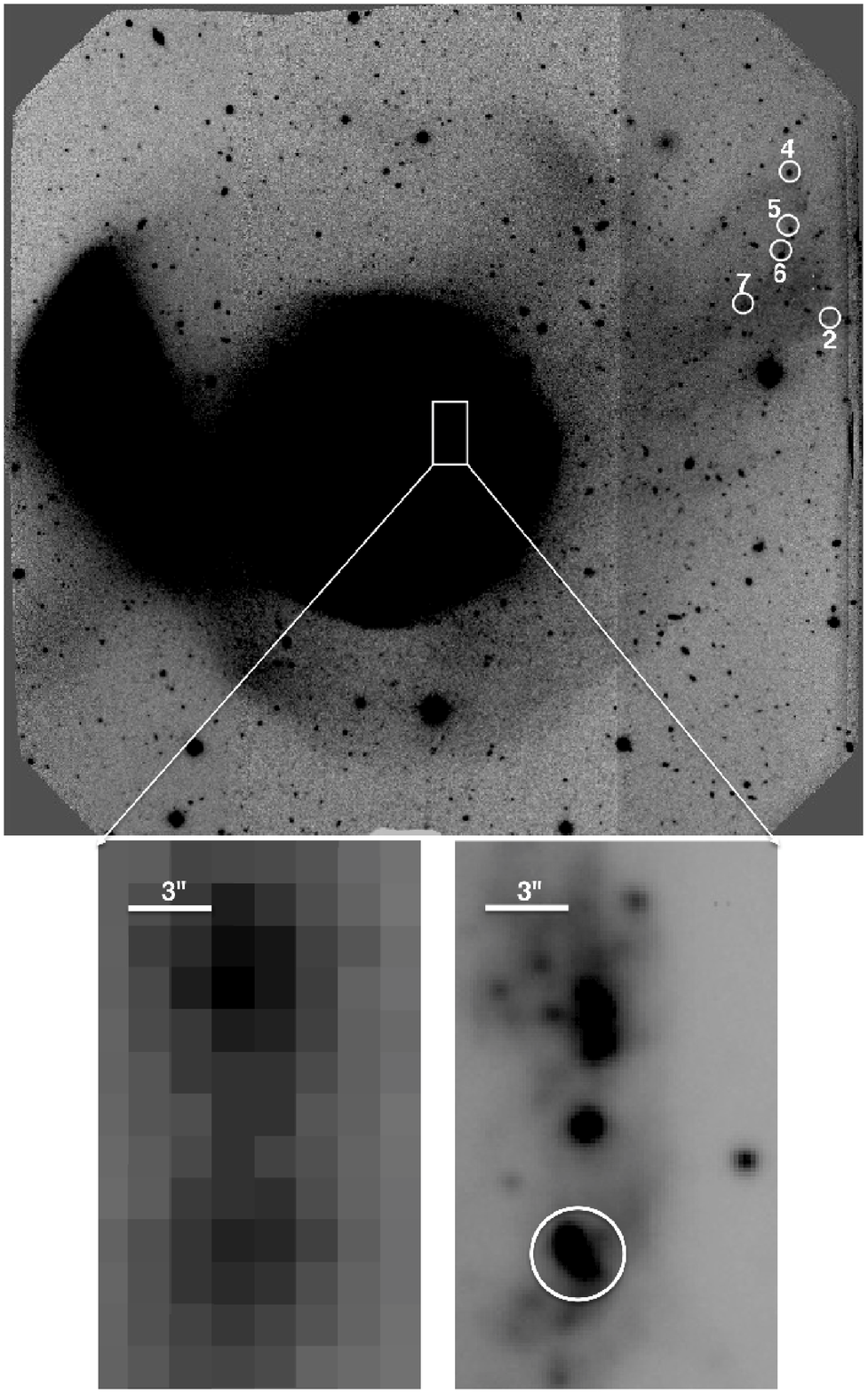}
\caption{r-band image of NGC 2782. UV-detected star-forming regions are marked with a white circle of 4$\arcsec$ radius. Shells structures can be seen around the main body of NGC 2782. The white square in the main body of NGC 2782 shows the location of the compact object with \textit{SDSS} spectrum. We also show a close-up of this object in the NUV and r-band filters (left and right images, respectively). The source is indicated by a white circle (RA=09:14:02.6, DEC=+40:06:47.2).}
\label{field2}
\end{figure}

We found 7 blue UV-emitting regions in the western tidal tail of NGC 2782. In Fig. \ref{fig1} we show NUV and r-band images of the gaseous tidal tail of NGC 2782. We also show a close-up of each region. In Fig. \ref{field2} we show an r-band image (taken with GEMINI/GMOS) of the main body of NGC 2782. Five of the seven detected regions are enclosed by a white circle (the other two regions are off the field of view). As already mentioned, we located the \textit{SDSS} spectrum of a compact region close to the center of NGC 2782. In Fig. \ref{field2} we show a zoom of such object.

\begin{figure}
\includegraphics[width=\columnwidth]{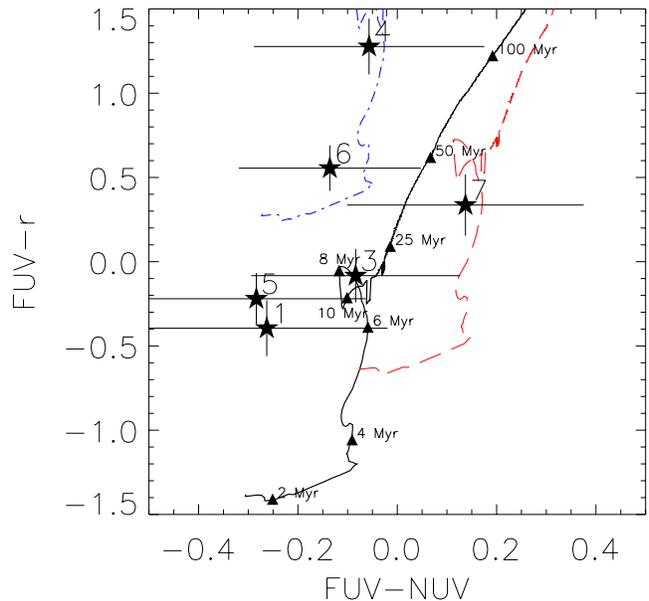}
\caption{\textit{GALEX} FUV-NUV versus FUV-r of the star-forming regions (black stars). Models from Starburst99 are shown as a solid black line (no extinction correction), dashed-dotted blue line (MW extinction law with a E(B-V)=0.30) and dashed red line (starburst extinction law from Calzetti et al. 1994 with a E(B-V)=0.30). Filled black triangles mark ages of 2, 4, 6, 8 , 10, 25, 50 and 100 Myrs.}
\label{fuv_nuv}
\end{figure}

\begin{figure}
\includegraphics[width=\columnwidth]{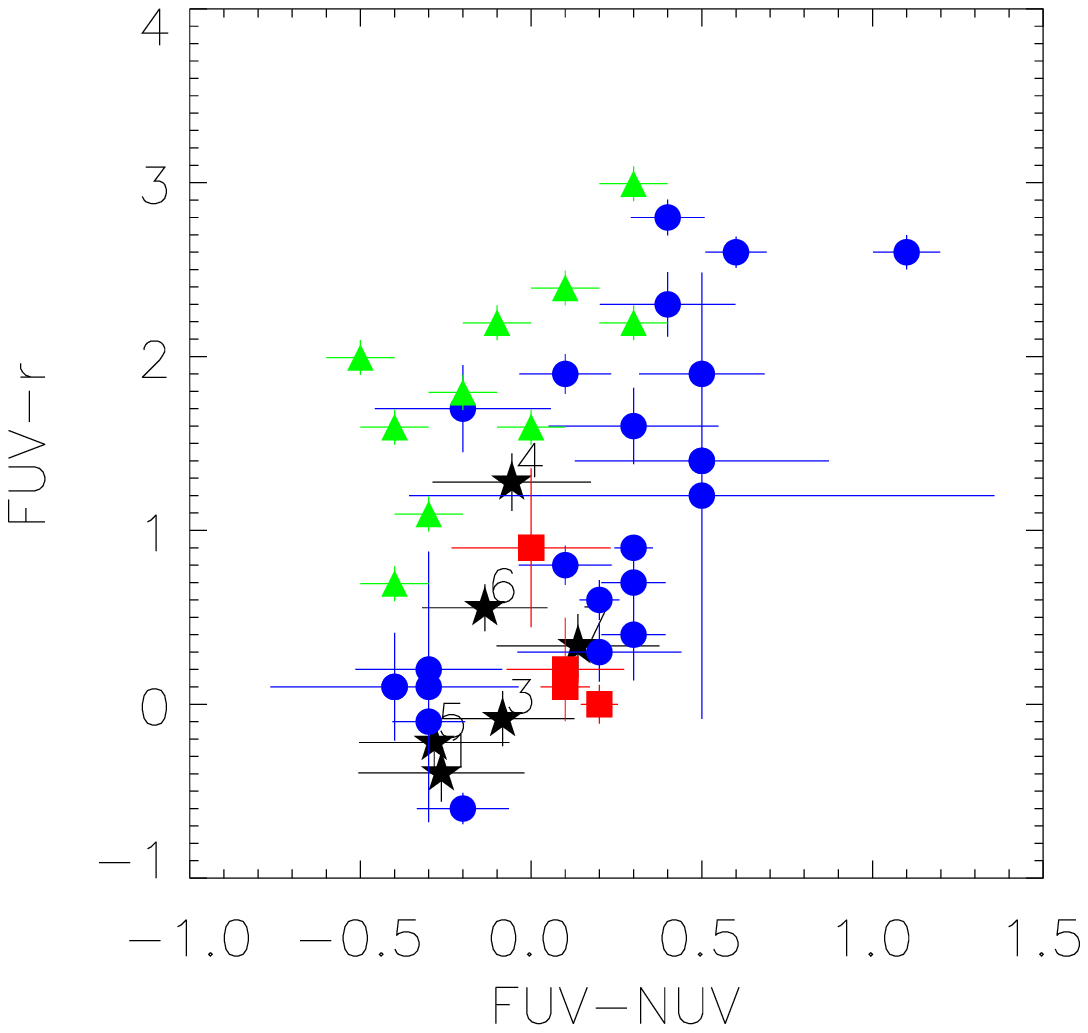}
\includegraphics[width=\columnwidth]{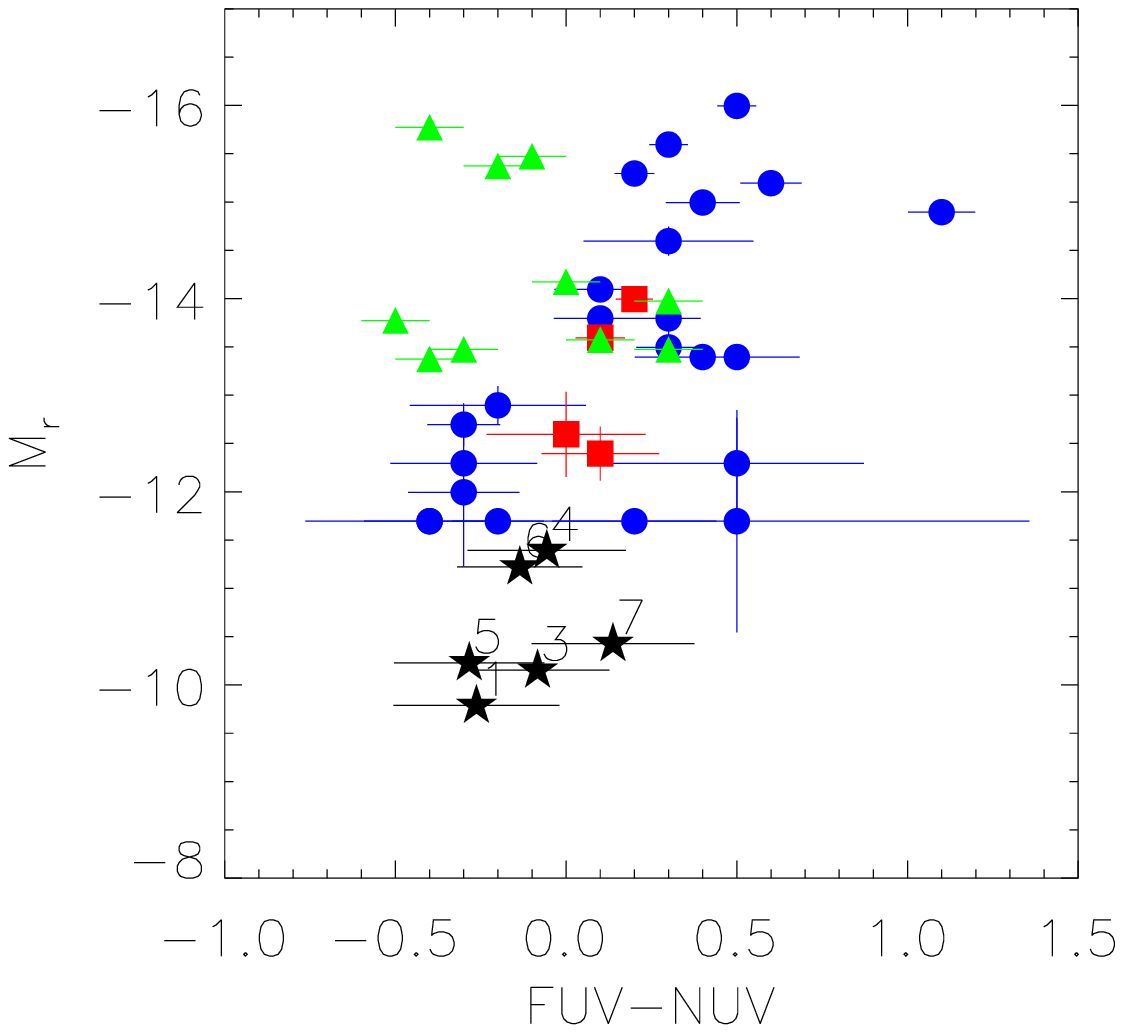}
\caption{Top Panel: \textit{GALEX} FUV-NUV versus FUV-r of the star-forming regions in NGC 2782 (black stars), Arp 82 (green triangles) and Arp 305 (blue circles and red squares). Bottom panel: r-band absolute magnitude versus FUV-NUV colors. The symbols are the same as in the top panel. The data for Arp 82 and Arp 305 regions were taken from Hancock et al. (2007,2009).}
\label{comparison}
\end{figure}

In Fig. \ref{field2} we show the r-band image of the main body of NGC 2782. This system clearly shows shells. From this figure, we note that in the region where the UV-emitting sources were found, there is diffuse emission, probably associated with an old stellar population. Diffuse emission is also present in the south and south-east regions of NGC 2782. 

In Figure \ref{fuv_nuv} we plot the FUV-NUV versus FUV-r colors for the regions located in the gaseous tidal tail of NGC 2782 (black stars). In the same plot, we include the SB99 models in the cases of no extinction (black line), MW (blue line) and starburst (red line) extinction laws. Best parameters (ages and respective uncertainties and extinctions) are listed in Table \ref{table1}. We found ages ranging from $\sim$1 to 11 Myr. Regions number 1, 5 and 6 are the youngest objects found in the gaseous tidal tail of NGC 2782. Interestingly, regions 5 and 6 are detected in the H$\alpha$ image of the system (de Mello et al. 2012, in preparation).

As described in \S3.3, we have calculated the color excess (E(B-V)) of the star-forming regions of NGC 2782. Regions 4, 6 and 7 have the highest color excess of this sample, with values of E(B-V)=0.26, 0.34 and 0.26, respectively. On the other hand, region 3 has the lowest reddening of 0.16 mag. In order to check these values, we have computed the color excess for each region by using the HI map of NGC 2782 (Smith 1994). In order to do that, we used equation C1 in Minchin et al. (2003) to estimate the HI column density and we assume the standard Galactic $N$(H)-to-extinction ratio of $N$(H)$/E(B-V)$= 5.8$\times$10$^{21}$ atoms cm$^{-2}$ mag$^{-1}$ (Bohlin et al. 1978). In Table \ref{table1} we listed the values of E(B-V) derived from neutral hydrogen. Within the uncertainties, the E(B-V) values from the two methods are reasonably consistent.

The mass estimates for the 6 star-forming objects (with photometrically derived ages) span a range of 3.9$<log M_{\odot}<$4.6 (Table \ref{table1}), which are similar to the masses inferred for the intergalactic H\,{\sc ii} regions detected in Mendes de Oliveira et al. (2004), for which the mean mass is log (M$_{\odot}$)=4.5. The mass of the star-forming regions in NGC 2782 are slightly lower than the masses of the regions found in the tail of Arp 285 (Smith et al. 2008), which range from 4.3$<log M_{\odot}<$5.9. In the same context, a few of the sources in the tail of Arp 82 (Hancock et al. 2007) have larger masses than those in NGC 2782, reaching masses of log (M$_{\odot}$)=7.9, in a similar way to the southern source located in the tail of Arp 105, which has a mass of log (M$_{\odot}$)=7.8 (Boquien et al. 2010). None of the regions detected in this work have masses typical of TDGs which are usually defined to have masses larger than $log (M_{\odot})\sim8.11$ (Duc \& Mirabel 1998).

We estimated the distance from the center of NGC 2782 to each individual region (Table \ref{table1}). The closest object is located at 26 kpc from the center of the galaxy, having an age of 6 Myr. If this object was thrown out of the disk of the main galaxy in some galaxy-galaxy interaction, it should have been expelled with an unlikely high velocity of $\sim$4200 km s$^{-1}$ (assuming a constant velocity, i.e. velocity=distance/age). As this object is one of the oldest ones (together with object 4) and also the closest one, it is similarly impossible for the other objects to have been thrown away from the disk of the main galaxy. This reasoning suggests these young objects were formed \textit{in situ}.

In Figure \ref{comparison} (top panel) we compared the FUV-NUV and FUV-r colors of the star-forming regions in NGC 2782 (black stars) with the regions detected in the tidal features of Arp 82 (green triangles) and Arp 305 (Hancock et al. 2007,2009). We used these systems as a comparison sample, given that both systems present tidal tails with FUV, NUV and R (or r) photometry, which allow us to do a fair comparison with the photometric properties of the star-forming regions in NGC 2782. In the case of Arp 82, R-band magnitudes were converted into r-band magnitudes by using the transformation given in Lupton (2005) and the g-r colors of the ``bridge TDG" regions of Arp 305 (given that there are no g-r colors for regions in Arp 82). In the case of Arp 305, regions belonging to tidal features are marked with a blue circle, while objects detected in the ``bridge TDG'' region are marked with red squares. In the bottom panel of Fig \ref{comparison} we compare the FUV-NUV color versus the r-band absolute magnitude for the same regions. From both plots, we can see that regions in NGC 2782 have similar colors to those displayed by the star-forming regions in Arp 82 and Arp 305 (although they tend to cluster at the blue end of the FUV-r range), but being less luminous.

\begin{table*}
\centering
\begin{minipage}[t]{\textwidth}
\scriptsize
\caption{Observed and Derived Properties for the UV detected regions}
\begin{tabular}{ccccccccccc}
\hline
System & R.A. & DEC. &  r\footnote{r-band magnitude inside a fixed aperture of 4" radius and corrected by MW extinction. Region 2 appears to have diffuse emission in the r-band image.} &  FUV\footnote{Fixed FUV-band magnitudes corrected by apertures (as described in \S 3.2) and MW extinction.} & FUV-r\footnote{FUV-r colors.} & FUV-NUV\footnote{Magnitudes were measured in a fixed aperture of 4" radius and corrected by MW extinction. FUV and NUV magnitudes were corrected by aperture, as described in \S3.2.}  &  E(B-V)\footnote{Left values: E(B-V) estimated as in \S 3.3. Right values: E(B-V) estimated by using the HI data of NGC 2782. Total extinction is the sum of the internal and Galactic extinctions (0.13 mag in this band).} & Age\footnote{Ages estimated from FUV-NUV and FUV-r. In parenthesis we show the ages estimated from the H${\beta}$ equivalent width and SB99 models.} & log M$_{\star}$ & Distance\footnote{Distance from the center of the galaxy, assuming a position angle and inclination of 0 degrees.} \\
   & 2000 &  2000 &  mag  & mag  & mag  & mag &  mag  &  Myr &  M$_{\odot}$  & kpc \\
\hline
1 & 09:13:48 & 40:09:56 & 22.87$\pm$0.04 &22.47$\pm$0.16 & -0.39$\pm$0.17  & -0.26$\pm$0.24   & 0.18/0.23      &1$\pm^{4}_{1}$        &  3.9 & 45\\ 
2 & 09:13:49 & 40:07:41 & $>$23                   &22.76$\pm$0.18 &...                            & -0.12$\pm$0.28    & .../0.23      &...                                     &  ...   & 31\\
3 & 09:13:50 & 40:10:20 & 22.50$\pm$0.03 &22.42$\pm$0.16 & -0.08$\pm$0.16 & -0.08$\pm$0.21    & 0.16/0.16 &5$\pm^{6}_{3}$            &  4.2 & 46\\
4 &09:13:51 & 40:08:39  & 21.27$\pm$0.01 &22.54$\pm$0.17 &  1.28$\pm$0.17  & -0.06$\pm$0.23   & 0.26/0.11 &11$\pm^{28}_{6}$         &  4.6  & 33\\
5 & 09:13:51 & 40:08:18 & 22.43$\pm$0.03 &22.21$\pm$0.15 & -0.22$\pm$0.15  & -0.28$\pm$0.22   & 0.22/0.21 &2$\pm^{4}_{1}$ ($\sim$4) &  3.9  & 31\\
6 & 09:13:51 & 40:08:08 & 21.44$\pm$0.01 &21.99$\pm$0.13 &  0.56$\pm$0.13  & -0.14$\pm$0.18   & 0.34/0.15 &3$\pm^{4}_{3}$ ($\sim$3)  &  4.0   & 30\\
7 & 09:13:52 & 40:07:47 & 22.23$\pm$0.03 &22.57$\pm$0.18 &  0.34$\pm$0.18  &  0.14$\pm$0.24   & 0.26/0.20 &6$\pm^{23}_{6}$ ($\sim$2) & 4.2    & 26\\
\hline
\vspace{-0.8cm}
\label{table1}
\end{tabular}
\end{minipage}
\end{table*}

\subsection{Main spectroscopic properties of the star-forming regions}
 
\begin{figure}
\includegraphics[width=\columnwidth]{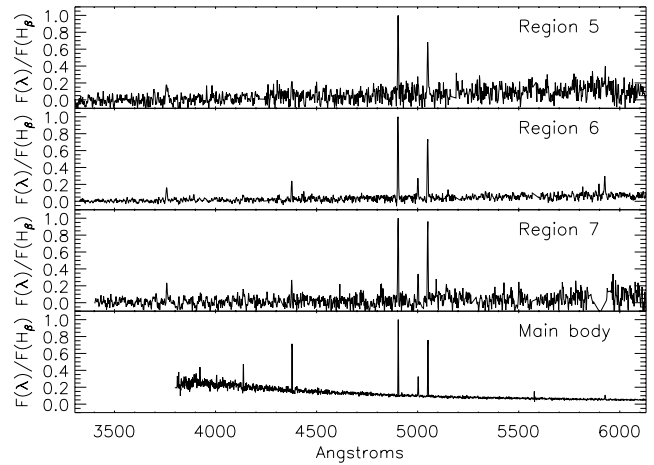}
\caption{GEMINI/GMOS B600 spectra of regions 5, 6 and 7 and \textit{SDSS} spectrum for the compact source close to the disk of NGC 2782 (bottom spectrum). Spectra were normalized to the peak intensity at the H$\beta$ line.}
\label{spec}
\end{figure}

In Fig.\ref{spec} (first three panels) we show the GMOS spectra (B600 grating) for the three regions located in the gaseous tidal tail of NGC 2782 for which spectroscopy was available. As expected from our selection criteria, these objects are very young. In the lowest panel of Fig. \ref{spec} we show the spectrum of the star-forming region located close to the main body of NGC 2782.

The three regions for which we have spectroscopy (5, 6 and 7), have radial velocities of 2556 km s$^{-1}$, 2531 km s$^{-1}$ and 2556 km s$^{-1}$, respectively, confirming that they are at the same distance as NGC 2782.

As described in \S3.4 we corrected the blue part of the spectra of regions 5, 6 and 7 by Galactic extinction using a value of E(B-V)=0.08. The results for the line ratios, all of them normalized to H$\beta$ are listed in Table \ref{table1b}.

\begin{table*}
\centering
\begin{minipage}[t]{\textwidth}
\scriptsize
\caption{Line intensities and oxygen abundances}
\begin{tabular}{ccccccccccc}
\hline
ID & [OII]\footnote{Line intensities relative to H$\beta$} & [OIII] &  [OIII] & H$\alpha$ & [NII] & EW$_{H\beta}$ &  E(B-V) & 12+log(O/H)\footnote{Following Pilyugin (2001a)}& 12+log(O/H)\footnote{Following Edmunds \& Pagel (1984). We have also used the method given in McCall et al. (1985) and Dopita \& Evans (1986), also finding similarly high values for the oxygen abundance.} &  12+log(O/H)\footnote{Following McGaugh (1991) and using a ionization parameter U=0.001 and M$_{u}$=60 M$_{\odot}$. For region with \textit{SDSS} spectrum (called MB in this table), this value was estimated using Denicol\'o, Terlevich \& Terlevich (2002).}  \\
  & 3727 & 4959 & 5007 &  6563 &  6584  & \AA  &    &   &    & \\
\hline
5 &  4.29$\pm$0.14 & 0.24$\pm$0.09 & 0.65$\pm$0.09  &...&...& 77   & 0.26\footnote{See \S3.4.} & 8.22$\pm$0.20 & 8.56$\pm$0.31 & 8.74$\pm0.20$\\
6 &  3.67$\pm$0.21 & 0.17$\pm$0.04 & 0.59$\pm$0.04  &...&...& 192 & 0.22$^{\textit{e}}$              & 8.31$\pm$0.20 & 8.65$\pm$0.31 & 8.81$\pm0.20$\\
7 &  3.75$\pm$1.07 & 0.34$\pm$0.08 & 0.73$\pm$0.09  &...&...& 286 & 0.34$^{\textit{e}}$              & 8.30$\pm$0.20 & 8.60$\pm$0.31 & 8.78$\pm0.20$\\
MB &             ...           & 0.28$\pm$0.01 & 0.73$\pm$0.02  & 2.01$\pm$0.05  & 0.57$\pm$0.01 &  33  & 0.85 & ... &...       & 8.72$\pm$0.10\\
\hline
\vspace{-0.8cm}
\label{table1b}
\end{tabular}
\end{minipage}
\end{table*}

Using the EW(H$\beta$) (single stellar population, Salpeter IMF and solar metallicity) we found ages of $\sim$4 Myr, $\sim$3 Myr and $\sim$2 Myr for regions 5, 6 and 7, respectively. For region 5 and 6, the spectroscopic estimate of the ages is in agreement with the age obtained from the colors, i.e. 2 and 3 Myr. In the case of region 7, the spectroscopic estimate of the age is lower than the age obtained from colors (6 Myr). These values are consistent within the uncertainties, however, we can not exclude an old stellar population co-existing with this star-forming region, similar to what was found in star-forming regions in the bridge of M81 and M82 (de Mello et al. 2008b).

Given that we are using semi-empirical and empirical calibrations in estimating the metallicities of the star-forming regions (and not direct estimates of these parameters), we adopt a fixed error of 0.2 dex in the case of the calibrations given by McGaugh (1991), as discussed in van Zee et al. (2006). We note that van Zee et al. (2006) found that oxygen abundances estimated by using the method shown in McGaugh (1991) could be over-predicted by 0.07$\pm$0.10 dex. In the case of the calibrations given by Edmunds and Pagel (1984) we adopt an error of 0.31 dex and in the case of Pilyugin (2001a), we used an error of 0.2 dex, as suggested by the author. 

For the region located close to the body of NGC 2782, with the \textit{SDSS} spectrum, we found an oxygen abundance of 12+log(O/H)=8.72$\pm$0.10 (using the N2 method, see Table \ref{table1b}). This value is closer to the upper branch of the R$_{23}$ calibration curve than to the lower branch. In view of this fact, we speculate that abundances in regions 5, 6 and 7 can be estimated by using the upper branch of the R$_{23}$ calibration. The oxygen abundances we derived for regions 5, 6 and 7 (using Pilyugin 2001a, Edmunds \& Pagel 1984 and McCall 1991, respectively) are listed in Table \ref{table1b}. The lowest oxygen abundance estimations were obtained by using the Pilyugin (2001a) approach, while the highest values were obtained from the McGaugh (1991) method. Using this latter method, we found that regions 5, 6 and 7 have abundances of 12+log(O/H)=8.74$\pm$0.20, 8.81$\pm$0.20 and 8.78$\pm$0.20, which are quite high values. It is interesting to note that if abundances in regions 5 and 6 were estimated by using the lower branch of the R$_{23}$ calibration (see Figure 9 in McGaugh 1991), their abundances would be at most 12+log(O/H)$\sim$7.8, 7.7 and 7.8, respectively, and these results would have been quite different than the abundances estimated for the region located on the main disk of NGC 2782. By using the same data set, Werk et al. (2011) found abundances of 12+log(O/H)=8.66, 8.72 and 8.56 for the same regions. Within the uncertainties, the abundances derived by us and by Werk et al. (2011) are consistent. In this case, our estimations of the oxygen abundances of the star-forming regions in NGC 2782 gives an additional measure of the uncertainty on the metallicities of these systems.

Taking into account the uncertainties, sources 5, 6 and 7 have similar oxygen abundances to those derived for the nuclear region of NGC 2782, i. e. 12+log(O/H)=8.63 (Moustakas \& Kennicutt 2006, as estimated by using the recipe given in Pettini \& Pagel 2004) and also are similar to the abundance determined by Engelbracht et al. (2008) for NGC 2782 (12+log(O/H)=8.59$\pm$0.10). By using the oxygen abundances of the regions listed above (plus two other objects), Werk et al. (2011) suggested a flat metallicity gradient for NGC 2782. Our analysis support this idea, however, new spectroscopic data for other H\,{\sc ii} regions along the western tidal tail of NGC 2782 are necessary to confirm this scenario.

The abundances of regions 5, 6 and 7 (listed in Table \ref{table1b}) are similar than the solar value given in Denicol\'o, Terlevich \& Terlevich (2002) and are higher (or of the same order) than the values found by Mendes de Oliveira et al. (2004), Duc \& Mirabel (1998) and Croxall et al. (2009) for intergalactic H\,{\sc ii} regions and TDGs in interacting systems.

\section{Discussion}

In this paper we have searched for star-forming regions in the gaseous tidal tail of the merger remnant NGC 2782. We used archival and new data to perform this analysis. Here, we summarize the main findings of our study:

\begin{enumerate}

\item We have identified 7 young UV-emitting sources in the HI tail of NGC 2782. Some of these regions are resolved in small clusters when high resolution optical images are used.

\item All these objects have young ages (1 Myr$<$Age$<$11 Myr) and masses ranging from  10$^{3.9}$ M$_{\odot}$ to 10$^{4.6}$ M$_{\odot}$.

\item Spectroscopic data for three of these objects confirm that they are at the same distance as NGC 2782. This results was also confirmed by Werk et al. (2011). We found that regions 5, 6 and 7 have high metallicities: 12+log(O/H)=8.74, 8.81 and 8.78, respectively. 

\end{enumerate}

Some of the main issues concerning star-formation outside galaxies are discussed below.

\subsection{The physical parameters of the intergalactic star-forming regions in NGC 2782}

In the past, a few authors have studied the properties of the stellar clusters in the gaseous tidal tail of NGC 2782. Using optical colors, Knierman (2007) found several blue clusters along the same tail. A few of our regions seem to be located at the same position as Knierman's clusters, as can be judged from their Fig. 1 (they did not publish coordinates). Using Fabry-Perot observations, Bournaud et al. (2004) reported just one H\,{\sc ii} region in the HI tail of NGC 2782. We detect an object in the same location of the Bournaud's H\,{\sc ii} region (region \#6 in our work), however, we found several other similar objects in the HI tail. Probably the detection limit of the Fabry-Perot observations of Bournaud et al. could not reveal all the star-forming regions in the HI tail. 

The fact that regions 5, 6 and 7 have quite high oxygen abundances suggests that these regions were formed from pre-enriched material. The material out of which these regions may have been formed could have been thrown out of the central region of the interacting galaxies that merged to form NGC 2782, forming the western gaseous tidal tail (scenario that is supported by N-body/SPH simulations, see Rupke et al. 2010). If star formation events already took place in this tidal tail, these could have polluted the intergalactic medium with metals and with an underlying old stellar population. Bastian et al. (2009) found that young stellar clusters can easily be destroyed within the first 10 Myr of their lives. Therefore, it remains plausible that the gaseous tidal tail of NGC 2782 already was a nursery of new stellar systems, which were destroyed by internal processes possibly through a phenomena called \textit{infant mortality} (see \S5.3), polluting the medium with metals. This fact, plus the distance to the parent galaxy and the young ages that these object present suggest that these star-forming systems were formed \textit{in situ}.

We note that the three spectroscopically detected regions in this work have similar ages to the star-forming regions found by Mendes de Oliveira et al. (2004), by construction, given that we chose objects of similar colors. However, they are different of those objects since they have higher oxygen abundances. On the other hand, regions 5, 6 and 7 of this work have oxygen abundances comparable to the values displayed by the TDG candidates in the M81 system (Croxall et al. 2009).

\subsection{Tidal tails: the birthplace of new stellar systems}

Galaxy-galaxy interactions are an effective mechanism to strip neutral gas out of galaxy disks, forming extended gaseous tidal tails (e.g. Hibbard et al. 2001). Recently, Mullan et al. (2011) used the Wide Field Planetary Camera 2 on the \textit{Hubble Space Telescope} to search for compact stellar structures in the tidal tails of a sample of 13 interacting galaxies, where NGC 2782 was included. These authors found an absence of cluster candidates in the western tidal tail of NGC 2782, however, by using \textit{GALEX} and Gemini data we have cataloged several star-forming region candidates, with ages $<$11 Myr, in the gaseous western tidal tail of this system. All of our sources were detected in regions where the HI column densities reach values above the threshold to trigger star formation (log\textit{N$_{HI}$}=20.6 cm$^{-2}$) found by Maybhate et al. (2007) in their study of cluster formation in tidal debris. 

In the western tidal tail of NGC 2782 no TDG was detected, however, several objects with IH\,{\sc ii} properties were found. Knierman et al. (2003) found a trend of formation of either TDGs or less massive systems but not both on tidal tails, for a sample of 4 mergers. In this context, the results found in this paper are in agreement with this previous work, given that NGC 2782 has only low mass systems.

\subsection{The fate of the star-forming regions in the HI tail of NGC 2782}

If these systems will become independent entities is not clear. It will depend on several parameters, such as the distance to the parent galaxies and total masses. In this context, Mengel et al. (2008) compare the dynamical masses for nine young clusters in the Antennae (6-9 Myr) with respect to the photometric masses. They found that almost all clusters in the Antennae are stable and bound. Internal processes, as ionizing radiation coming from OB stars, stellar winds and supernova could remove the interstellar medium that binds young star-forming regions (as stellar clusters) and allow these objects to become gravitationally unbound systems, in a phenomena called \textit{Infant Mortality} (Fall et al. 2005). Recently, Bastian et al. (2009) found that infant mortality is more important in the first 10 Myr of a cluster, contrary to what was proposed by Whitmore et al. (2007), who found that the disruption of a cluster takes about 100 Myr. In this context, the continuous formation of stellar clusters along the time, which are subsequent disrupted by the \textit{Infant Mortality} process, could increase the metallicity of a given region just due to stellar evolution. In the case of NGC 2782, the high oxygen abundance present at large distances from the main body of the system suggests that the intergalactic medium is already rich in metals. If the \textit{Infant Mortality} process already took place in this system, it could explain, at some level, the high metallicity observed in the gaseous tidal tail of NGC 2782. 

\subsection{Where is the molecular gas in merger galaxies?}

Although there is a good correlation between the location of UV sources with the presence of peaks in the HI distribution in tidal tails (e.g. Neff et al. 2005, de Mello et al. 2008a), it is well known that stars form in molecular clouds. As Knierman (2007) noted for NGC2782, it is an unexpected result to find star formation in regions without molecular gas. From our analysis, several UV sources are located in the same region of the blue clusters of Knierman, where HI has been detected. In view of this scenario, where is the molecular gas in merger systems? Using a sample of seven mergers, Bryant \& Scoville (1999) found that the molecular gas is concentrated in the core of the mergers. In a similar way, Yun \& Hibbard (2001) found that the CO emission in mergers is placed in the central 2 kpc radius. Studying a sample of TDGs, Braine et al. (2001) detected CO in six objects of their sample. Smith (1991) found an HI mass of 1.4x10$^{9}$M$_{\odot}$ in the northeast tail of NGC 2782. Interestingly, neither Smith et al. (1999) nor Braine et al. (2001) detect CO in the location of the UV-emitting regions in NGC 2782.

Therefore, the results shown in this paper imply that star formation took place in the tidal tail a few million years ago. The lack of CO detection can be explained if we take into account Heithausen et al. (2000) results using the same instrument (and beam size) as Braine et al. (2001). They detected two molecular complexes in the nearby galaxy NGC 3077. Complex \#2 has a molecular mass of M=4x10$^{6}$M$_{\odot}$. This mass and their respective CO flux at the distance of NGC 2782 (\textit{S}=0.32 Jy Km s$^{-1}$, using equation 4 of Braine et al. 2001) are below the values that Braine et al.
(2001) defined as \textit{no detection of molecular gas} for NGC 2782 (\textit{S}$\leq$0.5 Jy Km s$^{-1}$). In a similar way, Walter et al. (2006) studied the molecular gas in the tidal arms near to NGC 3077 using interferometric data. They found several H\,{\sc ii} regions over the HI distribution in this area and one of these regions (object A) is associated with a molecular cloud complex. If we take the flux of this region (\textit{S}=8.2 Jy Km s$^{-1}$), located at 3.6 Mpc, and carry it to the distance of the stellar clusters found in NGC 2782, this region has a flux of \textit{S}=0.09 Jy Km s$^{-1}$ and it could not be detected in CO, following Table 2 in Braine et al. (2001). Our detected regions seem to be small objects, with typical masses of stellar clusters, therefore their masses of molecular gas could be low enough to remain undetected by present observations at the distances of our targets. 

\section{Conclusions}
In this paper we found 7 young UV-emitting sources in the gaseous tidal tail of the system NGC 2782. All these regions have stellar masses and ages typical of IH\,{\sc ii}. They are located across the HI distribution and not specifically on HI peaks, as also was reported by Mendes de Oliveira et al. (2004) for the IH\,{\sc ii} in the Stephan's Quintet and by Walter et al. (2006) for NGC 3077. When an optical r-band image is taken into account, these UV sources can be resolved in several smaller systems. We found that three objects (region 5, 6 and 7) have high metallicities (12+log(O/H)=8.74$\pm$0.20, 8.81$\pm$0.20 and 8.78$\pm$0.20, respectively). This result could be explained if some high abundance gas from the center of the merging galaxies could be expelled, forming tidal tails. An additional contribution for the observed high abundances could be related to star formation events. In this case, the process known as \textit{infant mortality}, which predicts that stellar clusters are dissolved after $\sim$10 Myrs, could help explaining the observed high metallicities. Further spectroscopic data are necessary to constrain the membership of other detected regions in the vicinity of the interacting system, to study metallicities, to know how the intergalactic medium is enriched by this type of object and to compare dynamical and photometric masses, in order to know the fate of these systems.

\section*{acknowledgements}
We would like to thank the referee for the very useful comments that improved this paper considerably. We thank Gladys Vieira-Kober and Elysse Voyer for making the H$\alpha$ data of NGC 2782 available to us prior to publication. S. T--F. acknowledges the financial support of FONDECYT (Chile) through a post-doctoral position, under contract 3110087 and FAPESP through the Doctoral position, under contract 2007/07973-3. S. T--F. would also like to thank the NASA's Goodard Space Flight Center and The Catholic University of America for support during visit where part of this work was developed. C. M. d. O. acknowledges support from the Brazilian agencies FAPESP (projeto tem\'atico 2006/56213-9), CNPq and CAPES. D.FdM acknowledges support from \textit{GALEX} grant NNG06GG45G and ADP grant NNX09AC72G. S.S.J acknowledges FAPESP for the pos-doc grant 09/05181-8. F. U--V. acknowledges the financial support of FAPESP through the Master position, under contract 2007/06436-4. \textit{GALEX} is a NASA Small Explorer, launched in 2003 April. We gratefully acknowledge NASA's support for construction, operation, and science analysis for the \textit{GALEX} mission, developed in cooperation with the Centre National d' Etudes Spatiales of France and the Korean Ministry of Science and Technology.
This research has made use of the NASA/IPAC Extragalactic Database (NED) which is operated by the Jet Propulsion Laboratory, California Institute of Technology, 
under contract with the National Aeronautics and Space Administration.

\label{lastpage}
\end{document}